\documentclass{sigchi}

\toappear{\scriptsize Permission to make digital or hard copies of all or part of this work for personal or classroom use is granted without fee provided that copies are not made or distributed for profit or commercial advantage and that copies bear this notice and the full citation on the first page. Copyrights for components of this work owned by others than ACM must be honored. Abstracting with credit is permitted. To copy otherwise, or republish, to post on servers or to redistribute to lists, requires prior specific permission and/or a fee. Request permissions from Permissions@acm.org. \\ \\
{\emph{UIST '19}, October 20-23, 2019, New Orleans, LA, USA} \\
\copyright~2019 Association of Computing Machinery. \\
ACM ISBN 978-1-4503-6816-2/19/10\ ...\$15.00. \\
http://dx.doi.org/10.1145/3332165.3347873}

\usepackage{balance} 
\usepackage{graphics} 
\usepackage[T1]{fontenc} 
\usepackage{txfonts}
\usepackage{siunitx}
\usepackage{mathptmx}
\usepackage[pdflang={en-US},pdftex]{hyperref}
\usepackage{color}
\usepackage{booktabs}
\usepackage{textcomp}
\usepackage{amsmath}
\usepackage{algorithm}
\usepackage{multirow}
\usepackage[noend]{algpseudocode}

\usepackage{microtype} 
\usepackage{ccicons} 
\usepackage{todonotes}

\def\doctitle{StateLens: A Reverse Engineering Solution for Making Existing Dynamic Touchscreens Accessible}
\def\plaintitle{StateLens: A Reverse Engineering Solution for Making Existing Dynamic Touchscreens Accessible \vspace{-1pc}}
\def\plainauthor{Anhong Guo, Junhan Kong, Michael Rivera, Frank F. Xu, Jeffrey P. Bigham}

\def\plainkeywords{Reverse engineering; dynamic interfaces; touchscreen appliances; accessibility; crowdsourcing; computer vision; conversational agents.}

\makeatletter
\def\url@leostyle{%
 \@ifundefined{selectfont}{
 \def\UrlFont{\sf}
 }{
 \def\UrlFont{\small\bf\ttfamily}
 }}
\makeatother
\def\pprw{8.5in}
\def\pprh{11in}

\definecolor{linkColor}{RGB}{6,125,233}
\hypersetup{%
 pdftitle={\doctitle},
 pdfauthor={\plainauthor},
 pdfkeywords={\plainkeywords},
 pdfdisplaydoctitle=true, 
 bookmarksnumbered,
 pdfstartview={FitH},
 colorlinks,
 citecolor=black,
 filecolor=black,
 linkcolor=black,
 urlcolor=linkColor,
 breaklinks=true,
 hypertexnames=false
}

\newcommand{\thing}[1]
{\href{https://www.thingiverse.com/thing:#1}{thing:#1}}

\begin{document}

\title{\plaintitle}

\numberofauthors{1}
\author{
 \alignauthor{Anhong Guo, Junhan Kong, Michael Rivera, Frank F. Xu, Jeffrey P. Bigham} \\ \smallskip
 \affaddr{Human-Computer Interaction Institute, Carnegie Mellon University, Pittsburgh, PA, USA}\\
 \email{ \{anhongg, jbigham\}@cs.cmu.edu, \{junhank, mlrivera, fangzhex\}@andrew.cmu.edu }
}

\maketitle

\begin{abstract}
Blind people frequently encounter inaccessible dynamic touchscreens in their everyday lives that are difficult, frustrating, and often impossible to use independently. Touchscreens are often the only way to control everything from coffee machines and payment terminals, to subway ticket machines and in-flight entertainment systems. Interacting with dynamic touchscreens is difficult non-visually because the visual user interfaces change, interactions often occur over multiple different screens, and it is easy to accidentally trigger interface actions while exploring the screen. To solve these problems, we introduce {\em StateLens} \textemdash{} a three-part reverse engineering solution that makes existing dynamic touchscreens accessible. First, StateLens reverse engineers the underlying state diagrams of existing interfaces using point-of-view videos found online or taken by users using a hybrid crowd-computer vision pipeline. Second, using the state diagrams, StateLens automatically generates conversational agents to guide blind users through specifying the tasks that the interface can perform, allowing the StateLens iOS application to provide interactive guidance and feedback so that blind users can access the interface. Finally, a set of 3D-printed accessories enable blind people to explore capacitive touchscreens without the risk of triggering accidental touches on the interface. Our technical evaluation shows that StateLens can accurately reconstruct interfaces from stationary, hand-held, and web videos; and, a user study of the complete system demonstrates that StateLens successfully enables blind users to access otherwise inaccessible dynamic touchscreens.

\end{abstract}



\section{Introduction}

\begin{figure}[!t]
 \centering
 \vspace{0.5pc}
 \includegraphics[width=.95\linewidth]{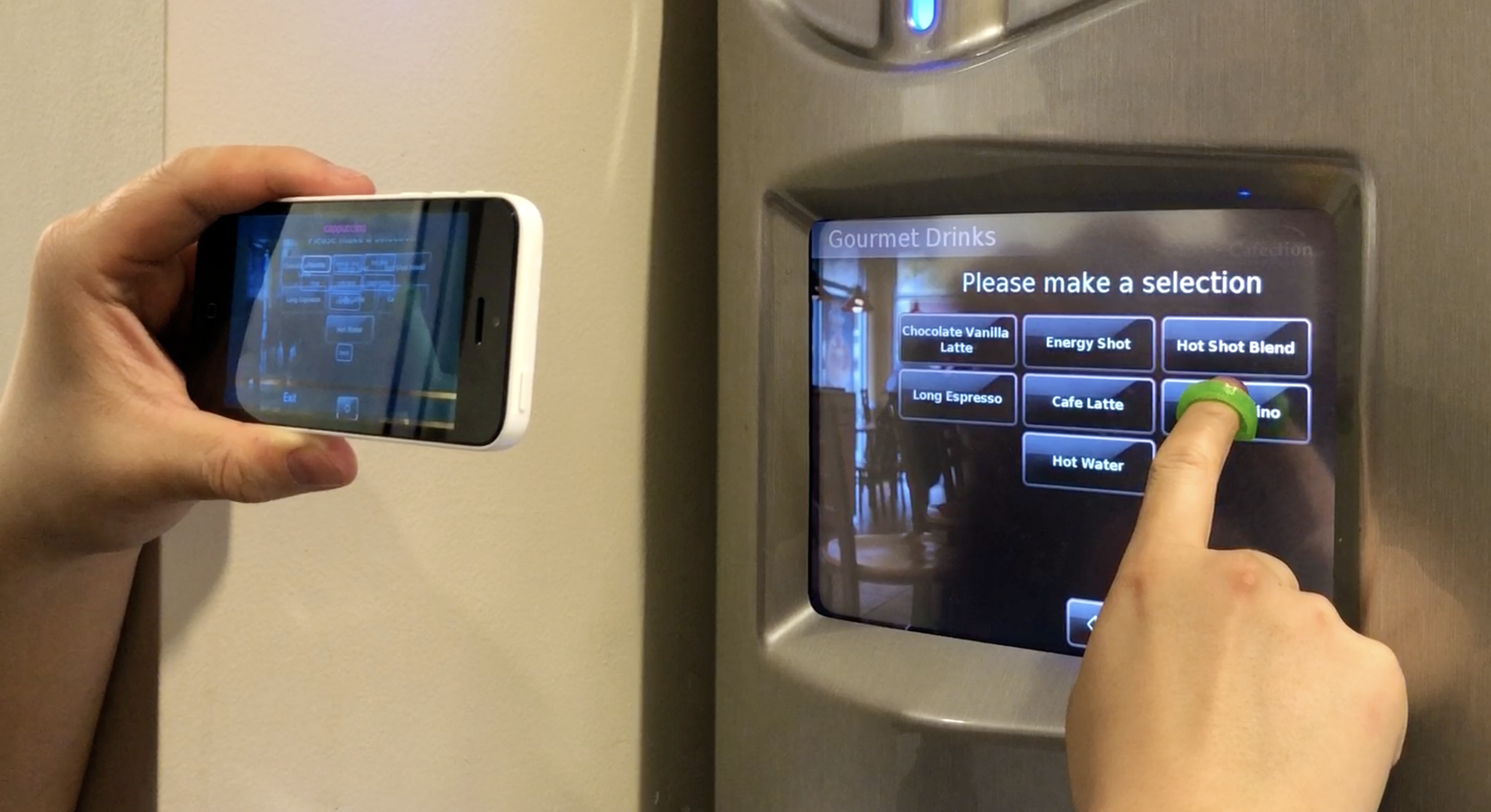}
 \caption{StateLens is a system that enables blind users to interact with touchscreen devices in the real world by {\em (i)} reverse engineering a structured model of the underlying interface, and {\em (ii)} using the model to provide interactive conversational and audio guidance to the user about how to use it. A set of 3D-printed accessories enable capacitive touchscreens to be used non-visually by preventing accidental touches on the interface.}
 \vspace{-1pc}
 \label{fig:userusingstatelens}
\end{figure}

\begin{figure*}[ht]
\centering
\vspace{0.5pc}
	\includegraphics[width=\textwidth]{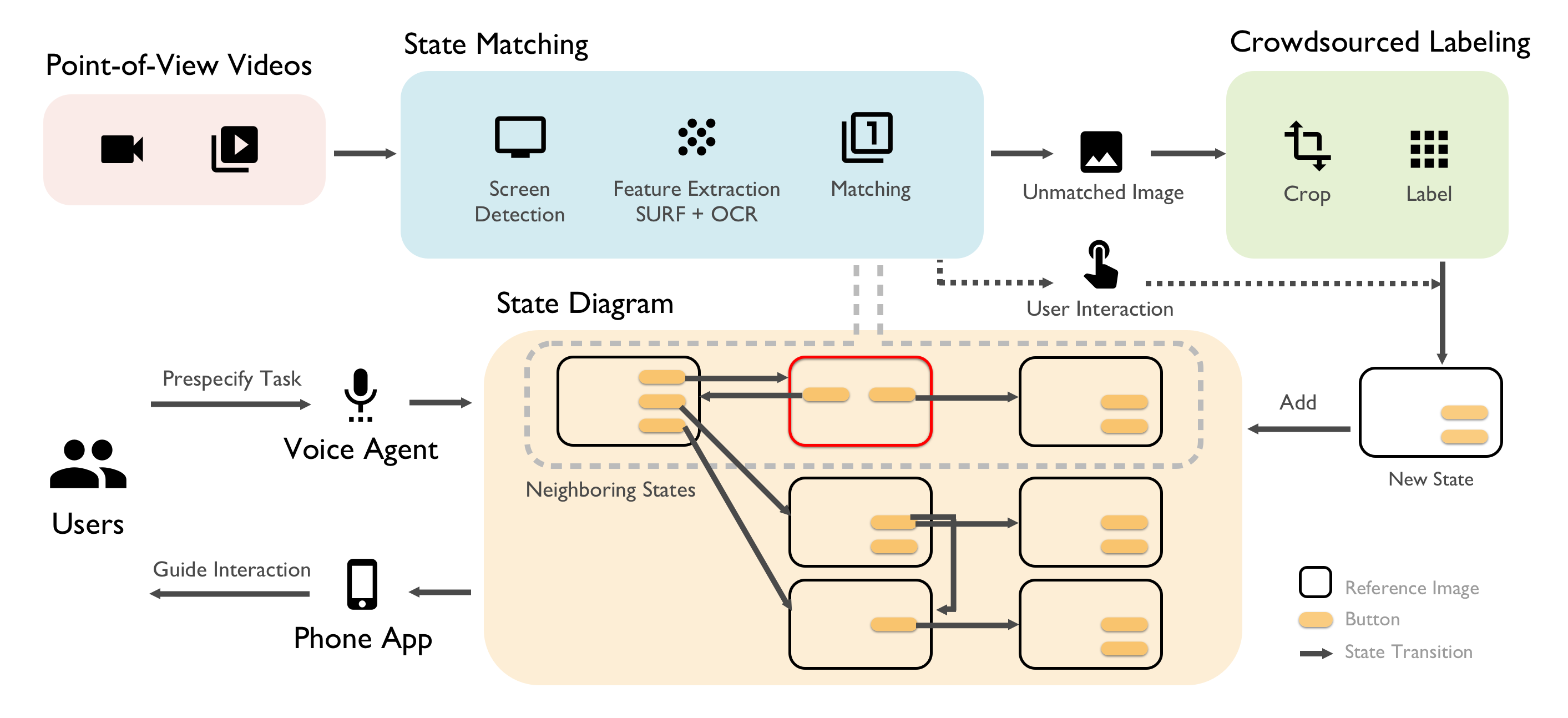}
 \caption{StateLens uses a hybrid crowd-computer vision pipeline to dynamically generate state diagrams about interface structures from point-of-view usage videos, and using the diagrams to provide interactive guidance and feedback to help blind users access the interfaces.}
 \vspace{-.5pc}
 \label{fig:System}
 \vspace{1pc}
\end{figure*}

Inaccessible touchscreen interfaces in the world represent a long-standing and frustrating problem for people who are blind. Imagine sitting down for a 12-hour flight only to realize that the entertainment center on the seatback in front of you can only be controlled by its inaccessible touchscreen; imagine checking out at the grocery store and being required to tell the cashier your pin number out loud because the checkout kiosk is an inaccessible touchscreen; and, imagine not being able to independently make yourself a coffee at your workplace because the fancy new coffee machine is controlled only by an inaccessible touchscreen. Such frustrating accessibility problems are commonplace and pervasive.

Making touchscreen interfaces accessible has been a long-standing challenge in accessibility \cite{fusco2014using,guo2016vizlens,clearspeech}, and some current platforms are quite accessible ({\em e.g.}, iOS). Solving all of the challenges represented by the combination of difficult issues for public touchscreen devices has remained elusive: {\em (i)} touchscreens are inherently visual so a blind person cannot read what they say or identify user interface components, {\em (ii)} a blind person cannot touch the touchscreen to explore without the risk of accidentally triggering something they did not intend, and, {\em (iii)} a blind person does not have the option to choose a different touchscreen platform that would be more accessible and cannot get access to the software or hardware to make it work better. This paper is about enabling blind people to use the touchscreens they encounter {\em in-the-wild}, despite the fact that nothing about how these systems are designed is intended for their use.

Most prior work on making touchscreens accessible has assumed access to change or add to the touchscreen hardware or software. For example, physical buttons were added to the side of the screen to provide a tactile way to provide input \cite{vanderheiden2011creating, vanderheiden2000flexible}. Slide Rule developed multi-touch gestures that could control touchscreens non-visually \cite{sliderule}, which have informed the popular VoiceOver screen reader on the iPhone. In the real world, users cannot control the touchscreens they encounter, and many are not accessible. In response, recent work has considered making existing interfaces accessible using computer vision and crowdsourcing to interpret the interfaces on-the-fly and provide immediate feedback to users \cite{guo2016vizlens}. This approach can work for many static interfaces, but struggles when the interface changes dynamically (as most touchscreens do), and cannot solve the problem of how a blind user could interact with a touchscreen without accidentally triggering touches.

This paper introduces {\em StateLens}, a reverse engineering solution for making existing dynamic touchscreens accessible. StateLens works by reverse engineering state diagrams of existing interfaces from point-of-view usage videos using a hybrid crowd-computer vision pipeline (Figure~\ref{fig:System}). Using the state diagrams, StateLens automatically generates conversational agents that guide blind users to prespecify tasks (Figure~\ref{fig:agentinteraction}). The StateLens iOS application then provides interactive guidance and feedback to help blind users access the interfaces (Figure~\ref{fig:userusingstatelens}). StateLens is the first system to enable access to dynamic touchscreens in-the-wild, that addresses the very hard case in which blind users encounter a touchscreen that is inaccessible and unfamiliar, which they cannot modify the hardware or software, and whose screen updates dynamically to show new information and interface components.

A known challenge for touchscreen interfaces is that they cannot easily be explored non-visually without the risk of accidentally triggering functions on the screen. Slide Rule developed the notion of ``risk-free exploration'' to counter this problem \cite{sliderule}, but their solution (requiring multiple taps instead of just one) requires being able to modify how the touchscreen operates. StateLens is intended to work on touchscreens already installed in the world that are not possible to be modified. To do this, we introduce a set of simple 3D-printed accessories that allow users to explore without touching the screen with their finger, and perform a gesture to activate touch at a desired position. These accessories add ``risk-free exploration'' to existing touchscreen devices without modifying the underlying hardware or software.

In a formative study, we first identified key challenges and design considerations for a system to provide access to dynamic touchscreen interfaces in the real world. Our technical evaluation showed that StateLens can accurately reconstruct interface structures from stationary, hand-held, and web usage videos. Furthermore, the generated state diagrams effectively reduced latency and prevented errors in the state detection process. Then through a user study with 14 blind participants, we showed that the conversational agent, the iOS application, and the 3D-printed accessories collectively helped blind users access otherwise inaccessible dynamic touchscreen devices effectively. StateLens represents an important step for solving this long-standing accessibility problem, and its technical approach may find applications broadly for augmenting how people interact with the touchscreens they encounter.

\section{Related Work}
Our work is related to prior work on {\em (i)} reverse engineering user interfaces, and {\em (ii)} improving the accessibility of existing physical interfaces. StateLens is intended to solve a long-standing and hard problem at the intersection of these spaces.

\subsection{Reverse Engineering User Interfaces}
A core feature of StateLens is its ability to reverse engineer user interfaces {\em in-the-wild} based on videos of their use. Substantial prior work exists in reverse engineering user interfaces using computer vision from ``pixels.'' This approach has been recognized as one of the most universally applicable methods for understanding a user interface's components, which is somewhat surprising given that at some level most user interfaces have been created with libraries that in some way had knowledge of their semantics. Unfortunately, that information is often either lost or inaccessible once the user interface makes it into a running system. StateLens is intended to make user interfaces accessible that are on public touchscreen devices, to which access is purposefully restricted. 

Prior work on reverse engineering of user interfaces has mainly used sceenshots or screencast videos. These approaches have looked to automatically extract GUI components from screenshot images in order to decouple GUI element representation from predefined image templates \cite{banovic2012waken, chang2011associating, dixon2010prefab, hurst2010automatically, yeh2009sikuli}, to augment existing interfaces through understanding of GUI components \cite{banovic2012waken, dixon2010prefab}, and to extract interaction flows from screencast videos and screen metadata \cite{kim2014crowdsourcing, li2017sugilite, li2010framewire, zhang2017interaction}. Prefab \cite{dixon2010prefab} identifies GUI elements using GUI-specific visual features, which enables overlaying advanced interaction techniques on top of existing interfaces. Sikuli \cite{yeh2009sikuli} uses computer vision to identify GUI components in screen captures for search and automation in the interfaces. Hurst {\em et al.} \cite{hurst2010automatically} combine a number of useful computer vision techniques with mouse information to automatically identify clickable targets in the interface. Chang {\em et al.} \cite{chang2011associating} propose an accessibility and pixel-based framework, which also allow for detecting text and arbitrary word blobs in user interfaces. Waken \cite{banovic2012waken} recognizes UI components and activities from screencast videos, without any prior knowledge of that application. 

Some of the prior work has gone beyond the task of identifying individual GUI components from static photos, and looked instead to extract interaction flows from screencast videos and screen metadata provided by the system API. For instance, FrameWire \cite{li2010framewire} automatically extracts interaction flows from video recordings of paper prototype user tests. Using Android's accessibility API, Sugilite \cite{li2017sugilite} and Interaction Proxies \cite{zhang2017interaction} extract the screen structures, in order to create automation and improve mobile application accessibility. Kim {\em et al.} \cite{kim2014crowdsourcing} apply a crowdsourcing workflow to extract step-by-step structure from existing online tutorial videos. 

StateLens builds on this rich literature, and applies a hybrid crowd-computer vision pipeline to automatically extract state diagrams about the underlying interface structures from point-of-view usage videos. In contrast to prior work, StateLens is a solution for reverse engineering existing physical interfaces through much noisier point-of-view videos rather than screenshots or prototyped GUIs.

\subsection{Improving Accessibility for Physical Interfaces}
Many physical interfaces in the real world are inaccessible to blind people, which has led to substantial prior work on systems for making them accessible. Many specialized computer vision systems have been built to help blind people read the LCD panels on appliances \cite{fusco2014using, clearspeech, tekin2011real}. These systems have tended to be fairly brittle, and have generally only targeted reading text and not actually using the interface.

Crowd-powered systems robustly make visual information accessible to blind people. VizWiz lets blind people take a picture, speak a question, and get answers back from the crowd within approximately 30 seconds \cite{bigham2010vizwiz}. More than 10,000 users have asked more than 100,000 questions using VizWiz \cite{vizwiz-dataset}. Users often ask questions about interfaces \cite{vizwiz-chi2013}, but it can be difficult to map the answers received, {\em e.g.}, ``the stop button is in the middle of the bottom row of buttons'', to actually using the interface because doing so requires locating the referenced object in space ({\em e.g.}, place a finger on the button).

Other systems provide more continuous support. For example, Chorus:View \cite{lasecki2013answering} pairs a user with a group of crowd workers using a managed dialogue and a shared video stream. ``Be My Eyes'' matches users to a single volunteer over a video stream \cite{bemyeyes}. These systems could more easily assist blind users with using an interface, but assisting in this way is likely to be cumbersome and slow. RegionSpeak \cite{zhong2015regionspeak} and Touch Cursor \cite{cursors} enable spatial exploration of the layout of objects in a photograph using a touchscreen. This can help users understand the relative positions of elements, but they still have the challenge of physically locating the elements in space on the real interface in order to use it.

Static physical interfaces can be augmented with tactile overlays to make them accessible. Past research has introduced fabrication techniques for retrofitting and improving the accessibility of physical interfaces. For example, RetroFab \cite{ramakers2016retrofab} is a design and fabrication environment that allows non-experts to retrofit physical interfaces, in order to increase usability and accessibility. Facade \cite{facade} is a crowdsourced fabrication pipeline to help blind people independently make physical interfaces accessible by adding a 3D-printed augmentation of tactile buttons overlaying the original panel.

VizLens \cite{guo2016vizlens} is a screen reader to help blind people use inaccessible static interfaces in the real world ({\em e.g.}, the buttons on a microwave). Our work goes beyond VizLens by enabling access to dynamic touchscreens. Without the 3D-printed accessories introduced in this paper, VizLens would not work for touchscreens. VizLens users would also need to take pictures when the screen changes, which is difficult. With VizLens, at each step, a good picture must be taken, labeled, and only afterwards can users explore the buttons on the single screen. Each screen iteration would take several minutes, making it cumbersome to use for dynamic interfaces.

VizLens::State Detection is able to do limited adaptation to dynamic interfaces by matching against every possible state and providing feedback based on the best match. However, because of changing display states and screen layouts, exploring and activating UI components across multiple screens is difficult (analogous to finding one's way in a new city). By generating and using state diagrams, StateLens enables a crucial interaction of previewing and prespecifying tasks through a conversational agent (analogous to using map applications to plan trips and follow turn-by-turn directions). The 3D-printed accessories make exploration possible by bringing risk-free exploration to touchscreens.

\section{Formative Study}
We conducted a formative study to identify the key challenges and design considerations for a system to provide access to dynamic touchscreen interfaces in the real world. We conducted semi-structured interviews with 16 blind people about their experiences and challenges with public touchscreen appliances, and their strategies for overcoming these challenges. Then using a Wizard-of-Oz approach, we asked two participants to try using a touchscreen coffee machine with verbal instructions given by the researchers. We extracted key insights that reflected participants' challenges and strategies, which we used in the design of StateLens.

\subsection{Design Considerations}
Participants remarked that interfaces are becoming much less accessible as flat touch pads and touchscreens replace physical buttons. Touchscreen appliances mentioned by participants were very diverse, and their interfaces differed in size, type of functions and number of buttons.

\subsubsection{Supporting Independence}
Participants often resorted to sighted help when accessing public touchscreen appliances, and raised serious privacy concerns when asking others (often strangers) to help with entering sensitive information, {\em e.g.}, using credit card machines to complete financial transactions, or using sign-in kiosks at pharmacies and doctors' offices. Participants also mentioned sighted people giving incorrect or incomplete information because of a lack of patience or experience helping blind people. Our solution should enable blind people to independently access touchscreen devices without needing sighted assistance.

\subsubsection{Reducing Cognitive Effort}
For unfamiliar dynamic touchscreen devices, the amount of time and cognitive effort needed for blind people to explore, understand, and activate functions became quite heavy. Participants noted that if it were for a one-time use, it would not be worthwhile to invest the time and effort to learn the interface, which is much easier for sighted people. Our solution should support more fluid interactions to reduce blind users' cognitive effort in exploring the interface layout and accessing functions on complex and unfamiliar touchscreen devices.

\subsubsection{Enabling Risk-Free Exploration}
Participants shared their concerns and fears of accidentally triggering functions on inaccessible touchscreens. For example, a participant mentioned that once in a few weeks she would accidentally hit the settings button on her fridge's touchscreen panel, then she needed to call someone to come and check on it, which has been a huge burden.

When attempting to use existing inaccessible touchscreen devices, participants found holding their fingers in mid-air while trying to explore and locate the buttons to be very awkward and unusable, which also often resulted in accidental touches. Therefore, our solution should support ``risk-free exploration'' to enable blind users freely explore without accidentally triggering functions on the screen.

\section{Risk-Free Exploration}
Risk-free exploration allows blind users to freely explore without accidentally triggering functions on the screen, all without modifying the underlying hardware or software of the device.

\subsection{Thingiverse Survey}
\begin{table*}[t] 
 \centering
 \renewcommand{\arraystretch}{1.3} 
 \begin{tabular}{c | p{12.5cm}}
 \textbf{Category}
 & \textbf{Example Thingiverse Items} 
 \\ \specialrule{.1em}{.05em}{.05em} 
 \newline
 Styluses (17)
 & iPad drawing pencils (\thing{8976}); capacitive stylus (\thing{2870398}, \thing{225001}); resistive stylus (\thing{1582974}, \thing{577056}); mouth sticks (\thing{1321021}); wrist-cuff stylus (\thing{1315004}); Nintendo 3DS Stylus (\thing{798010})
 \\ \hline
 Prosthetic Accessories (10)
 & prosthetic hands (\thing{1717809}, \thing{380665}, \thing{242639}); prosthetic finger (\thing{2527421}, \thing{2840850})
 \\ \hline
 Finger Caps (6)
 & thimbles around or over the fingertip (\thing{612664}, \thing{1044791}); thumb protectors (\thing{28722}); adapter to hold another object on finger (\thing{2133318})
 \\ \hline
 Buttons (4)
 & button grid for mouse input (\thing{2745606}); mechanical triggers for mobile phone games (\thing{2960274}); assistive button via phone's microphone input jack (\thing{1471760}); braille button input for phone (\thing{1049237})
 \\ \hline
 Joysticks (2)
 & touchscreen mounted capacitive joystick (\thing{2361676}, \thing{2361676})
 \vspace{0.5em}
 \end{tabular}
 \caption{Categorization of our Thingiverse survey results related to assistive technologies, touchscreens, and finger-based interactions. The number of items is shown next to each category name.}
 \label{tab:Thingiverse}
\end{table*}

We first conducted an exploratory search on Thingiverse to understand what openly available solutions exist for people to interact with touchscreens and see if they can enable risk-free exploration for blind people. We created a list of 11 search terms including: touchscreen accessibility; touchscreen stylus; screen stylus; capacitive screen input; resistive screen; input assistive; assistive finger cap; finger cap; 3D printed accessibility; conductive PLA accessibility; and prosthetic finger. These search terms resulted in a total of 103 existing designs. We then filtered results that were not related to accessibility or assistive technology ({\em{e.g.}}, raspberry pi and/or touchscreen cases), leading to a total of 39 relevant items.

Using an approach akin to affinity diagramming \cite{contextualDesign1997,chenReprise2016,hofmann2018greater}, we classified these items into five main categories of devices: styluses, prosthetic accessories, finger caps, buttons and joysticks. We show each of these categories with example items in Table~\ref{tab:Thingiverse}. Although the Thingiverse designs are closely related to assistive usage for touchscreens, none of them satisfy our need to enable blind users risk-free access to an existing touchscreen device. We used these categories to inspire design ideas for prototypes that take on familiar forms used in the Thingiverse accessibility community but also support risk-free exploration (Figure~\ref{fig:accessories}).

\subsection{Finger Ring Prototype}
Inspired by the finger cap designs from Thingiverse, we first created a 3D-printed ring that allows users to explore without touching the screen, and tilt their finger forward to perform a touch at a desired position (Figure~\ref{fig:accessories}A-C).

We tested this design in a pilot study with two blind participants (one female, age 48; one male, age 57). While the 3D-printed finger ring enabled our participants to explore without accidental triggers, participants also identified issues related with the design and suggested other solutions. For example, the location of the ring on the finger may vary for different users and different sessions during use, thus changing the actual position of touch. Furthermore, when pressing the finger and finger ring on the touchscreen, it was uncomfortable for the participants for certain angles and postures. This is worsened when they are asked to only use one finger to interact with the interfaces, in order to prevent accidental touches. 

\subsection{Design Variations}
\begin{figure}[b!] 
 \centering
 \vspace{-1pc}
 \includegraphics[width=0.48\textwidth]{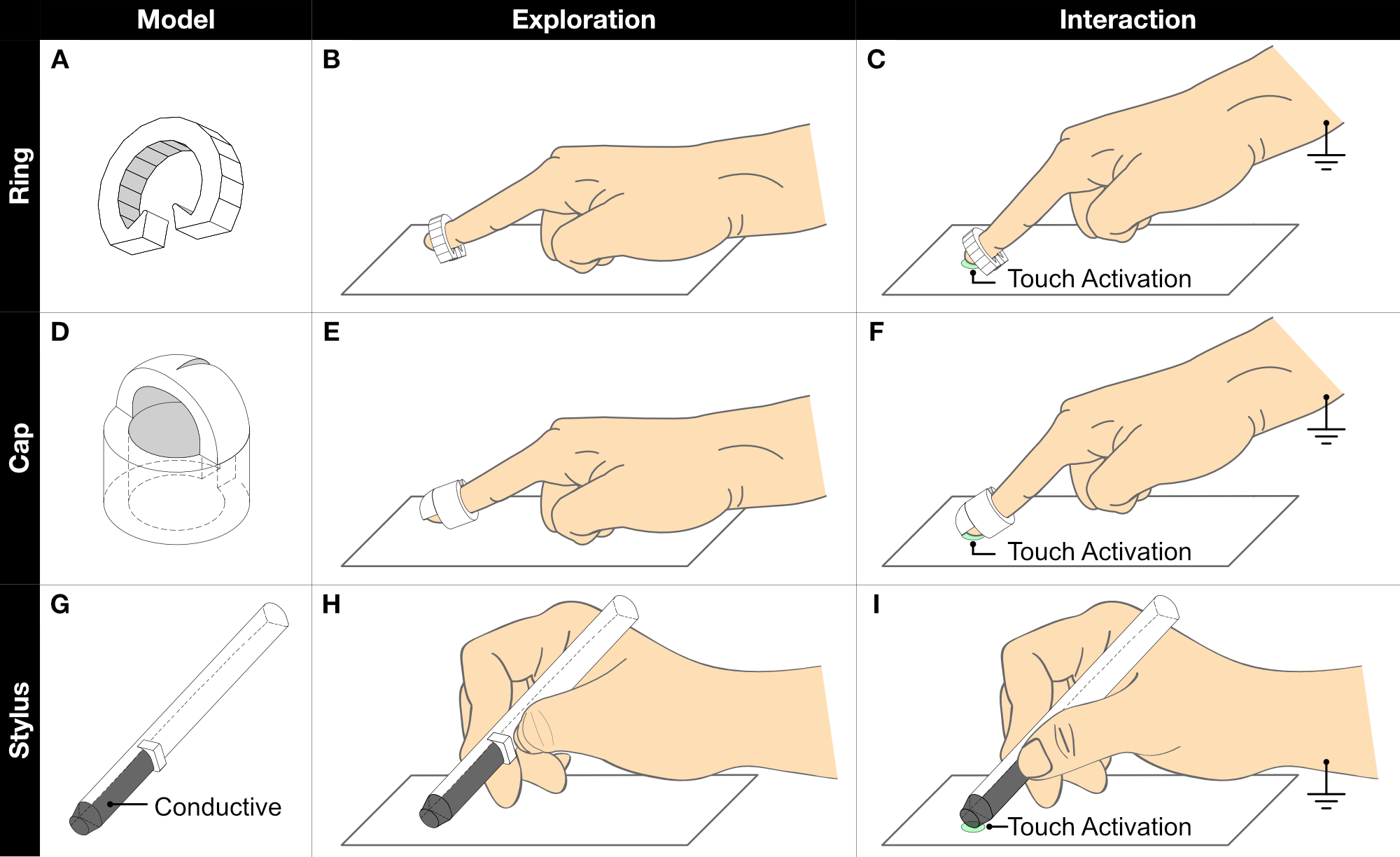}
 \caption{A set of 3D-printed accessories that prevent the wearer from accidentally triggering touches while exploring the interface. When desired, the wearer can activate a touch using either a tilt motion (B-C and E-F) or by touching a conductive trace on the accessory with a finger (H-I). These accessories elegantly add ``risk-free exploration'' to existing capacitive touchscreen devices without modifying the underlying hardware or software, which has been a major hurdle for past efforts. 3D models of these accessories are available at: \url{https://github.com/mriveralee/statelens-3dprints}}
 \label{fig:accessories}
\end{figure}

Informed by the participants' feedback to our initial prototype, we designed variations of 3D-printed accessories (Figure~\ref{fig:accessories}DG) that focus on improving stability and comfort during use. The designs aim to reduce the change of ``touchpoint'' when the user moves from exploration to interaction ({\em{i.e.}}, touch activation), and maintain consistency across sessions. We also focused on capacitive touchscreens rather than resistive touchscreens, since resistive screens usually require some pressure to activate so the issue of accidental activation is not as severe compared to capacitive touchscreens.

Our design variations consist of a finger cap (Figure~\ref{fig:accessories}D) and a conductive stylus (Figure~\ref{fig:accessories}G). The finger cap prevents accidental touches by shielding undesirable areas of the finger from touching the touchscreen. The cap has an opening on the finger pad that allows the user to tilt their finger to activate a touch (Figure~\ref{fig:accessories}DEF). Compared to the ring design, the finger cap's enlarged shielding area and top cover prevent accidental touches more effectively and ensure consistency across sessions. This finger-worn design also incorporates a slit so that when 3D printed with a flexible material ({\em e.g.}, thermoplastic polyurethane -- TPU), it can fit around fingers of different sizes. The stylus uses a conductive trace to trigger touches at the tip of the stylus when touched by a finger (Figure~\ref{fig:accessories}GHI). It provides a physical affordance to prevent accidental touches, by delineating the conductive and non-conductive regions with a rectangular bumper located on the side of the stylus. Conductive traces can be applied using conductive paint or printed with conductive PLA on a dual extrusion 3D printer. We had success with both techniques, though conductive PLA was more durable, while conductive paint can come off after repeated use.

\begin{figure*}[t]
 \centering
 \includegraphics[width=.98\textwidth]{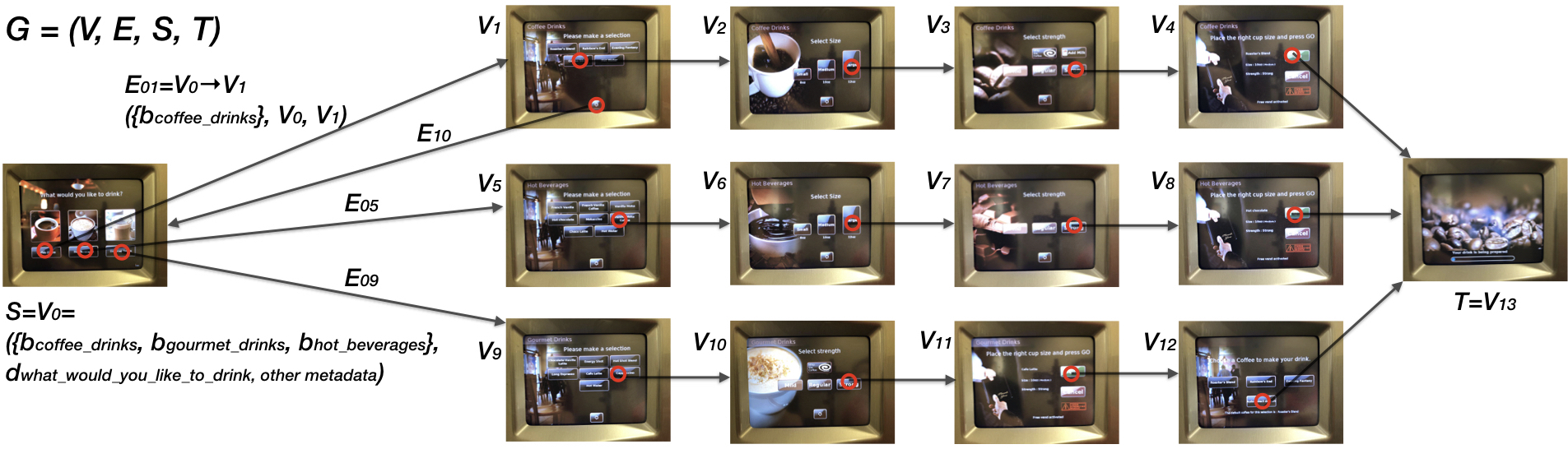}
 \caption{Visualization of how StateLens represents the coffee machine interface structure as a state diagram. Note only some edges are shown.}
 \label{fig:CV}
\end{figure*}

\section{StateLens}
StateLens uses a hybrid crowd-computer vision pipeline to dynamically generate state diagrams about interface structures from point-of-view usage videos, and to provide interactive feedback and guidance to help blind users access the interfaces through these diagrams. We use the coffee machine in Figure~\ref{fig:CV} as a running example.

\subsection{Generating the State Diagram}
The architecture of StateLens to generate state diagrams (Figure~\ref{fig:System}) involves capturing point-of-view usage videos from a variety of sources, representing state diagrams, detecting screen regions, identifying existing and new states, soliciting labels from the crowd, as well as recognizing user interactions.

\subsubsection{Capturing Point-of-View Usage Video}
StateLens takes point-of-view usage videos of dynamic interfaces from various sources as input to build up state diagrams about interface structures. These videos can be collected in many ways, including through existing IoT and surveillance cameras, through motivating sighted volunteers to contribute videos using mobile and wearable cameras, by encouraging manufacturers to share videos as a low-cost way to make their systems accessible to more people, and by mining existing demo and tutorial videos in online repositories. For example, a search on YouTube for ``coca cola freestyle machine demo'' produces many usage videos. In the current work, we demonstrate StateLens with videos captured from stationary cameras, hand-held mobile phones and web video repositories.

\subsubsection{Representing State Diagram}
StateLens represents the interface structure with a state diagram, as shown in Figure~\ref{fig:System} and the instantiation of the coffee machine shown in Figure~\ref{fig:CV}. We represent a state diagram as a directed graph $G=(V, E, S, T)$ where $S$ is the start state and $T=\{T_1, T_2, ..., T_n\}$ contains the end states where tasks are accomplished. Each node (state) $V_i \in V$ can be represented as $V_i = (\{b_1, b_2, ..., b_n\}, \text{descriptions, coordinates, other metadata})$, where $b_n$ is one of the interactive elements ({\em e.g.}, buttons) in state $V_i$. Each edge (transition) from state $V_i$ to state $V_j$ is $E_{ij} \in E$ that can be represented as $E_{ij} = (\{b_1, b_2, ..., b_m\}, V_i, V_j)$. Note that here $b_m$ represents the button identifier in the metadata of \textit{``from state''} that caused the state transition into \textit{``to state.''} Following our running example, the transition from the initial state $S = V_0 = (\{b_\text{coffee\_drinks}, b_\text{gourmet\_drinks}, b_\text{hot\_beverages}\}, \text{other metadata})$ to the coffee drink type state $V_1$ can be represented as: $E_{01} = V_0 \to V_1 = (\{b_\text{coffee\_drinks}\}, V_0, V_1)$, stating that by interacting with button ``Coffee Drinks'' in the initial state, we could get to the desired state for coffee drinks type selection. Similarly, the transition to go back to the initial state can be represented as: $E_{10} = V_1 \to V_0 = (\{b_\text{back}\}, V_1, V_0)$.

\subsubsection{Detecting the Screen}
StateLens detects whether a screen is present and its bounding box in the camera's field of view to filter out irrelevant video frames and random background content. Since there is no existing models for detecting touchscreen interfaces, we re-purpose state-of-the-art object detection models' output for this task. Using the Amazon Rekognition Object Detection API \cite{amazonrekognition}, StateLens first detects bounding boxes of object categories related to electronics and machines. If such bounding boxes exist and their sizes are above 10\% of the image size (aiming to filter out objects that are not the one of interest), StateLens crops the image using the bounding box to remove background noises for further processing. If not, StateLens checks whether the output labels with high confidence scores (above 55\%) appears in the above categories. If so, the full video frame is retained and used for further processing. If not, the frame is determined irrelevant and discarded. StateLens is quite lenient in this step to prevent accidentally removing relevant frames, in order to maintain a high recall.

\subsubsection{Identifying Existing States}
StateLens extracts two kinds of features and intelligently combines them (Figure~\ref{fig:System}): SURF (Speeded-Up Robust Features) \cite{bay2006surf} and OCR. StateLens first uses SURF feature detectors to compute key points and feature vectors in both the existing state reference images (Figure~\ref{fig:CV}) and the input image. The feature vectors are then matched using brute-force matcher with normalization type of L2 norms, which is the preferable choice for SURF descriptors. By filtering matches and finding the perspective transformation \cite{criminisi1999plane} between the reference-input image pairs using RANSAC (Random Sample Consensus) \cite{Fischler:1981}, StateLens is able to compute the ratios of inliers to the number of good matches for each existing state. It then uses the reference state with the distinctly highest ratio as the candidate matched state.

If the highest matched ratio across existing reference images is not high enough, meaning the match using only SURF features is not so confident, StateLens then uses the Google Cloud Vision API \cite{googlecloudvision} to compute OCR results for the input image and compares to the pre-computed OCR results of the state reference image. Similarity is defined as the ratio of longest common sequence (LCS) edit distance to the length of the OCR output results, and if above a threshold, the candidate matched state is finalized as the matched state. For example, matching $V_1$ against $V_5$ results in low confidence with SURF, then with additional information provided by OCR, StateLens is able to differentiate them. On the other hand, if both the matched inlier ratio and the OCR similarity score are below a certain threshold, StateLens determines it as not a match.

\subsubsection{Adding New States}
When a transition happens on the dynamic interface, the new state might not have been seen before. If an input image is not a match with the existing states, StateLens adds it to a candidate pool. Then, for the next images which are also added into this pool, they are matched against the existing candidates. Using this candidate pool approach, only when the same image is seen continuously across multiple frames, StateLens is confident enough to register it as a new state. Among the candidates identified as the same state, StateLens automatically selects the last one added to the pool as the reference image for this new state. We do so because the first few candidates often include transition residuals from the previous state, such as animations. We use a time window of 1 second for this process. On the other hand, if continuous unmatched states in the pool do not reach the window size to qualify as a new state, they are considered noise and the candidate pool will be cleared. Once a new state is registered, StateLens then sends it to the crowdsourced labeling pipeline to acquire more information such as the interface region, interaction components, and description (Figure~\ref{fig:System}).

\subsubsection{Soliciting Labels from the Crowd}
StateLens builds upon the crowdsourcing workflow in VizLens \cite{guo2016vizlens}, and uses a two-step workflow to label the area of the image that contains the interface assisted with screen detection results, and then label the individual interaction components assisted with OCR output (Figure~\ref{fig:System}). Crowd workers are first asked to rate the image quality, segment the interface region (with the generated screen bounding box as a start when available), indicate the approximate number of interaction components, and additionally provide a description of the interface state. Results are combined using majority vote. 

Crowd workers are then instructed to provide labels to the individual interaction components ({\em e.g.}, buttons) assisted with OCR output. Rather than requiring crowd workers to draw bounding boxes around all buttons and provide text annotations, the OCR-assisted crowd interface allows them to simply confirm or reject OCR-generated labels, and revise any errors. In this step, crowd workers also work in parallel, and the worker interface shows labeled elements to other workers as they are completed.

\subsubsection{Recognizing User Interaction}
Finally, StateLens captures the interaction component that triggered a state transition, {\em e.g.}, a button $b_n$ that contributes to the transition $E_{ij} = V_i \to V_j = (\{b_\text{n}\}, V_i, V_j)$. Essentially, StateLens uses the last image of the previous state $V_i$ before the state transition, transforms the input image to the reference image frame through warping, and detects the touchpoint location using skin color thresholding and other standard image processing techniques \cite{vezhnevets2003survey}. 

In the next section of Accessing the State Diagram, using the user interaction information, StateLens predicts the state that the interface could be transitioning to, and reduces the processing latency and errors by narrowing down the search space. Furthermore, StateLens aggregates these interaction traces to provide ranked usage suggestions to assist novice users. Note that recognizing finger touchpoint locations in naturalistic usage videos is not always possible or accurate, such as under extreme lighting conditions, or when users are wearing gloves. In those cases, StateLens will fallback to only using the state transition without the detailed interaction component as the triggering event, {\em e.g.}, $E_{ij} = V_i \to V_j = (\emptyset, V_i, V_j)$.

\subsection{Accessing the State Diagram}
To help blind users access the dynamic interfaces, StateLens takes advantage of the state diagram to efficiently identify states, integrates natural language agents, and interactively provides feedback and guidance (Figure \ref{fig:userusingstatelens}).

\subsubsection{Identifying States Efficiently and Robustly}
StateLens employs several techniques to enable efficient searching of states to reduce latency and prevent errors. First, when available, StateLens utilizes user's fingertip location to infer from the state diagram about the state that the interface has transitioned to, {\em e.g.}, using the button that the finger was on. Second, StateLens searches the neighbors of previously identified state for the best match, in case when the inferred state from the fingertip location matches poorly with input image. Third, in case the matching results with neighbor states are poor, StateLens gradually expands the search space to other states of the interface according to the distance, calculated as the shortest path in the state diagram. Fourth, StateLens applies a similar approach to the candidate pool for smoothing, and only when a new state has been seen continuously across multiple frames, it is confident enough to determine a state transition. Finally, the reference images can be pre-computed once in advance to improve processing speed. These techniques effectively reduces the search space, speeds up the state detection process, and improves the robustness of state detection, which we will validate in technical evaluation. Note that for performance reasons, only SURF features are used when detecting states to provide real-time feedback for blind users. This is because the screen detection and OCR processes have longer delays (\textasciitilde{1 second}). However, in the future, these processes can be sped up and the produced bounding boxes can be tracked across frames to offer better performance.

\begin{figure}[b!]
 \centering
 \includegraphics[width=.98\linewidth]{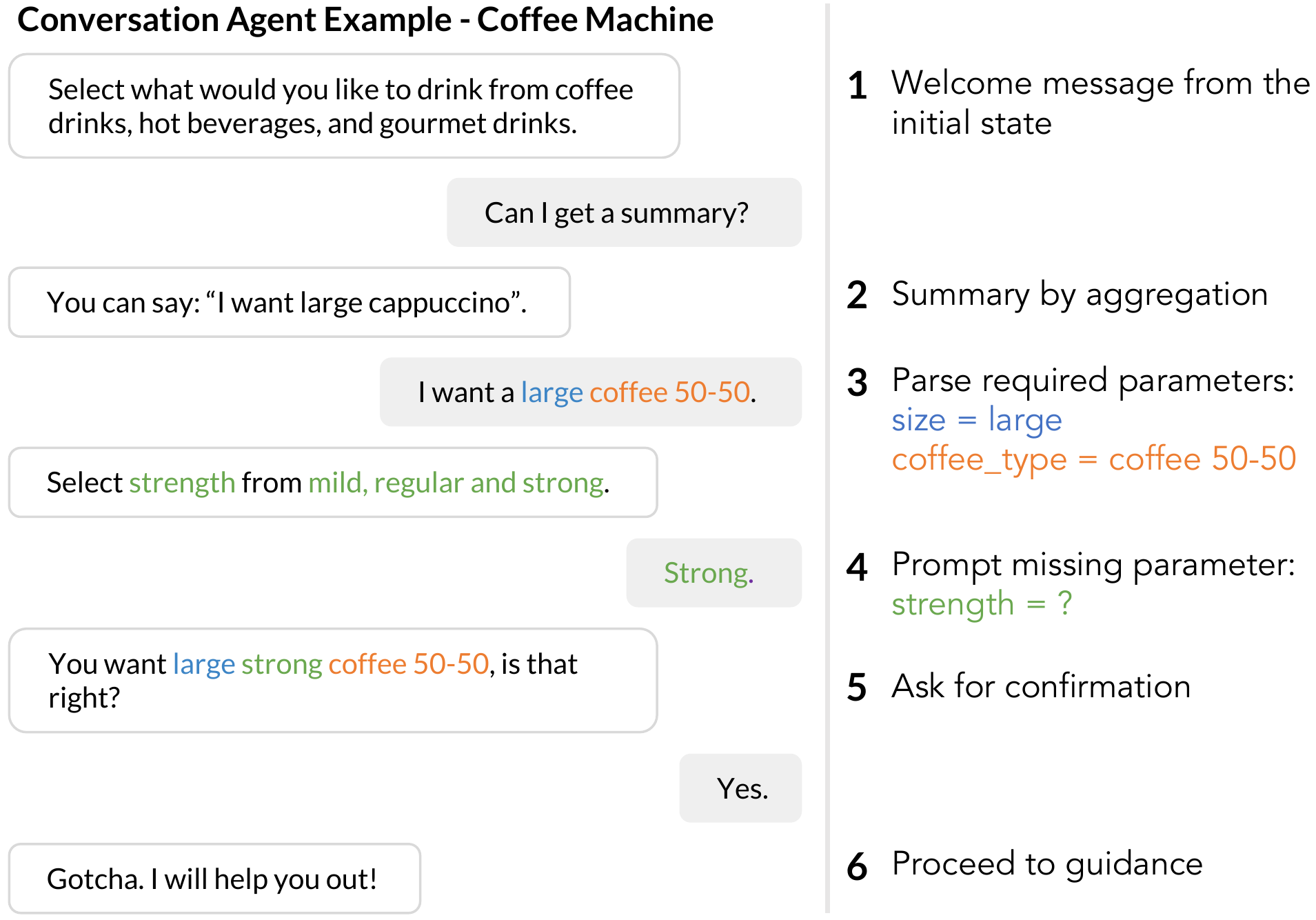}
 \caption{Sample interactions between a user and the coffee machine natural language conversational agent StateLens automatically generated.}
 \label{fig:agentinteraction}
 \vspace{-0.5pc}
\end{figure}

\begin{figure*}[t!]
\centering
	\includegraphics[width=.98\textwidth]{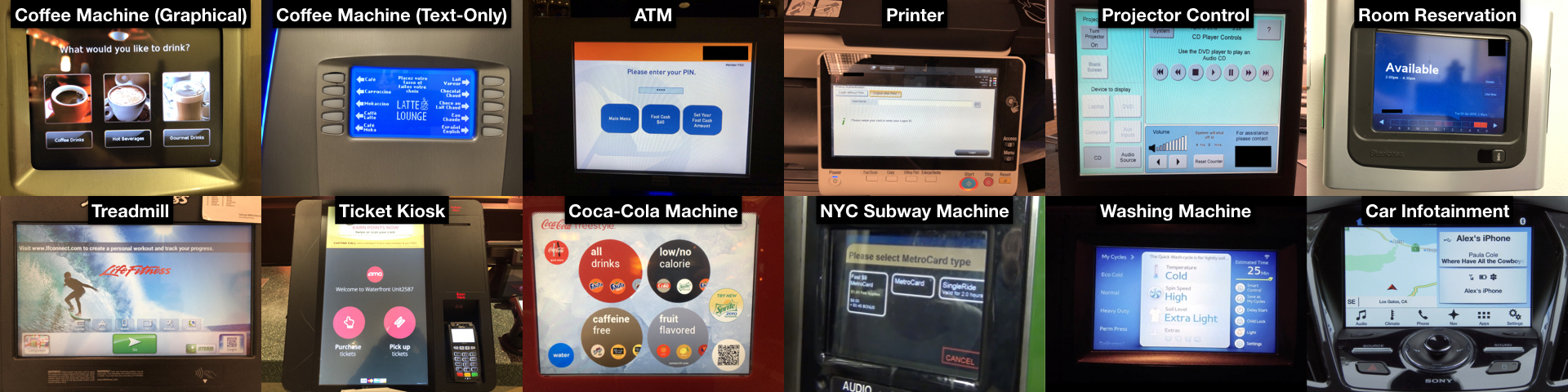}
 \caption{We evaluated how well StateLens reconstructs state diagrams from point-of-view usage videos across a wide range of interfaces, including an ATM, coffee machine (both graphical and text only), printer, projector control, room reservation panel, treadmill, ticket kiosk, Coca-Cola machine, subway ticket machine, washer, and car infotainment system.}
 \label{fig:interfaces}
\end{figure*}

\subsubsection{Enabling Natural Language Queries}
StateLens allows users to interact with a natural language conversational agent to prespecify the task they want to achieve. Inspired by our formative study, the goal of the conversational agent is to reduce the time and effort of the blind users to explore, understand, and activate functions on inaccessible and unfamiliar touchscreen interfaces. To do this, StateLens transforms all the possible paths (interaction traces) from $S$ to $T$ in the generated state diagram into different \textit{intents} ({\em e.g.}, to make coffee drinks, to make gourmet drinks), and the interactive element values in the edges $E_i$ along the path into required \textit{entities} for the intent and their attributes/values ({\em e.g.}, size: large/medium/small). Using the Google Dialogflow API \cite{googledialogflow}, StateLens automatically creates an agent for each device using these mappings. StateLens uses the description text from state $S$ as the welcome prompt and adds confirmation prompts at the end of intents. StateLens heuristically generates training samples for the intents and prompts to the required entities from the descriptive texts along different paths aforementioned. Because Dialogflow only requires a small number of user utterance samples for training, StateLens uses a random sample of entity values and concatenates with phrases such as ``Select ...'' to create training sentences. The created agent then guides the user through each required parameter needed to complete an interaction trace. Once all required entities are fulfilled, the StateLens iOS application will proceed to guiding the users to activate each button on the predefined interaction trace. A sample user-agent interaction is shown in Figure~\ref{fig:agentinteraction}.

\subsubsection{Generating Natural Language Summary}
StateLens uses the state diagram and the associated aggregation of interaction traces to automatically generate a natural language summary of the devices' popular use cases. This is designed to assist novice users get familiar with the device. To do this, StateLens ranks the aggregated interaction traces, then generates prompts for each trace based on the involved state and transition metadata as well as the corresponding interaction components. StateLens uses simple heuristic template-based generation methods that concatenate words like ``I want ...'' with most frequently selected button options, {\em i.e.} entities, as well as the descriptive text of the intent. This natural language summary is also integrated in the conversational agent (Figure~\ref{fig:agentinteraction}), and users can simply ask, {\em e.g.}, ``tell me a summary.''

\subsubsection{Providing Interactive Feedback and Guidance}
StateLens identifies the current state of the dynamic interface, and recognizes the user's touchpoint location to provide real-time feedback and guidance for blind users through the iOS application. For blind users accessing the interface with a 3D-printed accessory, a color marker on the accessory can be used to identify the touchpoint location. To make sure the touchpoint does not change from exploration to activation ({\em i.e.}, the problems Slide Rule \cite{sliderule} addressed with split tap, and VizLens \cite{guo2016vizlens} addressed with shifting the interaction point), we measured the ground truth touchpoint location and placed the color marker on the accessory accordingly.

StateLens then looks up the coordinates of the touchpoint in the current state's labeled interaction components, and announces feedback and guidance to the blind user, {\em e.g.}, ``state: coffee drinks, select strength; target: regular'', ``move up'', ``move left slowly'' and ``at regular, press it.'' StateLens also provides feedback to users when the interface is partially out of frame by detecting whether the corners of the interface are inside the camera frame. If not, it provides feedback such as ``move phone to right.'' Similarly, it provides feedback when it does not detect the interface or does not see a finger (using words or earcons \cite{blattner1989earcons} for ``no object'' or ``no finger'').

\begin{table*}
 \renewcommand{\arraystretch}{1.3}
\centering
\scriptsize
\begin{tabular}{c|c|c|c|c|c|c|c|c|c|c|c|c|c|c|c|c|c|c|c|c}
\textbf{} & \multicolumn{4}{c|}{\textbf{Coffee Machine (G)}} & \multicolumn{4}{c|}{\textbf{Coffee Machine (T)}} & \multicolumn{4}{c|}{\textbf{ATM}} & \multicolumn{4}{c|}{\textbf{Printer}} & \multicolumn{4}{c}{\textbf{Projector Control}} \\ 
\textbf{} & \multicolumn{2}{c|}{\textbf{Stationary}} & \multicolumn{2}{c|}{\textbf{Hand-held}} & \multicolumn{2}{c|}{\textbf{Stationary}} & \multicolumn{2}{c|}{\textbf{Hand-held}} & \multicolumn{2}{c|}{\textbf{Stationary}} & \multicolumn{2}{c|}{\textbf{Hand-held}} & \multicolumn{2}{c|}{\textbf{Stationary}} & \multicolumn{2}{c|}{\textbf{Hand-held}} & \multicolumn{2}{c|}{\textbf{Stationary}} & \multicolumn{2}{c}{\textbf{Hand-held}} \\ 
\specialrule{.1em}{.05em}{.05em} 
\textbf{\# of states} & \multicolumn{2}{c|}{14} & \multicolumn{2}{c|}{13} & \multicolumn{2}{c|}{11} & \multicolumn{2}{c|}{10} & \multicolumn{2}{c|}{11} & \multicolumn{2}{c|}{12} & \multicolumn{2}{c|}{10} & \multicolumn{2}{c|}{10} & \multicolumn{2}{c|}{13} & \multicolumn{2}{c}{9} \\ \hline
\textbf{\# of frames} & \multicolumn{2}{c|}{4,980} & \multicolumn{2}{c|}{2,580} & \multicolumn{2}{c|}{3,060} & \multicolumn{2}{c|}{2,310} & \multicolumn{2}{c|}{2,910} & \multicolumn{2}{c|}{2,340} & \multicolumn{2}{c|}{1,380} & \multicolumn{2}{c|}{1,980} & \multicolumn{2}{c|}{3,540} & \multicolumn{2}{c}{1,530} \\ \hline
\multirow{2}{*}{\textbf{SURF}} & 0.47 & 0.57 & 0.67 & 0.62 & 0.88 & 0.64 & 0.78 & 0.70 & 0.85 & 1.00 & 0.52 & 1.00 & 1.00 & 0.10 & 0.82 & 0.90 & 1.00 & 0.46 & 0.50 & 0.56 \\ \cline{2-21} 
 & \multicolumn{2}{c|}{0.52} & \multicolumn{2}{c|}{0.64} & \multicolumn{2}{c|}{0.74} & \multicolumn{2}{c|}{0.74} & \multicolumn{2}{c|}{0.92} & \multicolumn{2}{c|}{0.69} & \multicolumn{2}{c|}{0.18} & \multicolumn{2}{c|}{0.86} & \multicolumn{2}{c|}{0.63} & \multicolumn{2}{c}{0.53} \\ \hline
\multirow{2}{*}{\textbf{\begin{tabular}[c]{@{}c@{}}SD\\ +SURF\end{tabular}}} & 0.58 & 0.50 & 0.73 & 0.62 & 1.00 & 0.64 & 0.80 & 0.80 & 0.69 & 1.00 & 0.63 & 1.00 & 1.00 & 0.20 & 0.82 & 0.90 & 0.86 & 0.46 & 0.63 & 0.56 \\ \cline{2-21} 
 & \multicolumn{2}{c|}{0.54} & \multicolumn{2}{c|}{0.67} & \multicolumn{2}{c|}{0.78} & \multicolumn{2}{c|}{0.80} & \multicolumn{2}{c|}{0.81} & \multicolumn{2}{c|}{0.77} & \multicolumn{2}{c|}{0.33} & \multicolumn{2}{c|}{\textbf{0.86}} & \multicolumn{2}{c|}{0.60} & \multicolumn{2}{c}{0.59} \\ \hline
\multirow{2}{*}{\textbf{\begin{tabular}[c]{@{}c@{}}SURF\\ +OCR\end{tabular}}} & 0.72 & 0.93 & 0.65 & 0.85 & 0.73 & 1.00 & 0.67 & 1.00 & 1.00 & 1.00 & 0.75 & 1.00 & 1.00 & 0.40 & 0.63 & 1.00 & 1.00 & 0.54 & 0.67 & 0.67 \\ \cline{2-21} 
 & \multicolumn{2}{c|}{0.81} & \multicolumn{2}{c|}{0.73} & \multicolumn{2}{c|}{0.85} & \multicolumn{2}{c|}{0.80} & \multicolumn{2}{c|}{1.00} & \multicolumn{2}{c|}{0.86} & \multicolumn{2}{c|}{0.57} & \multicolumn{2}{c|}{0.77} & \multicolumn{2}{c|}{0.70} & \multicolumn{2}{c}{0.67} \\ \hline
\multirow{2}{*}{\textbf{\begin{tabular}[c]{@{}c@{}}SD+SURF\\ +OCR\end{tabular}}} & 1.00 & 0.93 & 0.85 & 0.85 & 0.91 & 0.91 & 0.77 & 1.00 & 1.00 & 1.00 & 0.75 & 1.00 & 0.86 & 0.60 & 0.63 & 1.00 & 1.00 & 0.62 & 0.70 & 0.78 \\ \cline{2-21} 
 & \multicolumn{2}{c|}{\textbf{0.96}} & \multicolumn{2}{c|}{\textbf{0.85}} & \multicolumn{2}{c|}{\textbf{0.91}} & \multicolumn{2}{c|}{\textbf{0.87}} & \multicolumn{2}{c|}{\textbf{1.00}} & \multicolumn{2}{c|}{\textbf{0.86}} & \multicolumn{2}{c|}{\textbf{0.71}} & \multicolumn{2}{c|}{0.77} & \multicolumn{2}{c|}{\textbf{0.76}} & \multicolumn{2}{c}{\textbf{0.74}} \\\specialrule{.2em}{.em}{.1em}
 \textbf{} & \multicolumn{4}{c|}{\textbf{Room Reservation}} & \multicolumn{4}{c|}{\textbf{Treadmill}} & \multicolumn{4}{c|}{\textbf{Ticket Kiosk}} & \multicolumn{2}{c|}{\href{https://youtu.be/Dm2VQHJih8Y}{\textbf{Coca-Cola}}} &
 \multicolumn{2}{c|}{\href{https://youtu.be/DcP5W9K-1E8}{\textbf{Subway}}} &
 \multicolumn{2}{c|}{\href{https://youtu.be/RLwrfBAMNAA}{\textbf{Washer}}} &
 \multicolumn{2}{c}{\href{https://youtu.be/Qr2bvXZRqyc}{\textbf{Car Control}}} \\ 
\textbf{} & \multicolumn{2}{c|}{\textbf{Stationary}} & \multicolumn{2}{c|}{\textbf{Hand-held}} & \multicolumn{2}{c|}{\textbf{Stationary}} & \multicolumn{2}{c|}{\textbf{Hand-held}} & \multicolumn{2}{c|}{\textbf{Stationary}} & \multicolumn{2}{c|}{\textbf{Hand-held}} & \multicolumn{2}{c|}{\textbf{Web}} & \multicolumn{2}{c|}{\textbf{Web}} & \multicolumn{2}{c|}{\textbf{Web}} & \multicolumn{2}{c}{\textbf{Web}} \\ 
\specialrule{.1em}{.05em}{.05em} 
\textbf{\# of states} & \multicolumn{2}{c|}{7} & \multicolumn{2}{c|}{8} & \multicolumn{2}{c|}{10} & \multicolumn{2}{c|}{10} & \multicolumn{2}{c|}{11} & \multicolumn{2}{c|}{14} & \multicolumn{2}{c|}{9} & \multicolumn{2}{c|}{16} & \multicolumn{2}{c|}{11} & \multicolumn{2}{c}{24} \\ \hline
\textbf{\# of frames} & \multicolumn{2}{c|}{1,560} & \multicolumn{2}{c|}{1,260} & \multicolumn{2}{c|}{1,260} & \multicolumn{2}{c|}{4,500} & \multicolumn{2}{c|}{1,470} & \multicolumn{2}{c|}{3,480} & \multicolumn{2}{c|}{2,100} & \multicolumn{2}{c|}{6,630} & \multicolumn{2}{c|}{5,940} & \multicolumn{2}{c}{17,940} \\ \hline
\multirow{2}{*}{\textbf{SURF}} & 0.83 & 0.71 & 0.33 & 1.00 & 1.00 & 0.10 & 1.00 & 0.80 & 0.47 & 0.73 & 0.53 & 0.57 & 0.23 & 1.00 & 0.48 & 0.94 & 0.46 & 0.55 & 0.79 & 0.46 \\ \cline{2-21} 
 & \multicolumn{2}{c|}{0.77} & \multicolumn{2}{c|}{0.50} & \multicolumn{2}{c|}{0.18} & \multicolumn{2}{c|}{\textbf{0.89}} & \multicolumn{2}{c|}{0.57} & \multicolumn{2}{c|}{0.55} & \multicolumn{2}{c|}{0.37} & \multicolumn{2}{c|}{0.64} & \multicolumn{2}{c|}{0.50} & \multicolumn{2}{c}{0.58} \\ \hline
\multirow{2}{*}{\textbf{\begin{tabular}[c]{@{}c@{}}SD\\ +SURF\end{tabular}}} & 0.86 & 0.86 & 0.50 & 0.75 & 1.00 & 0.10 & 1.00 & 0.50 & 0.58 & 0.64 & 0.64 & 0.50 & 0.33 & 0.67 & 0.71 & 0.94 & 0.78 & 0.64 & 0.73 & 0.67 \\ \cline{2-21} 
 & \multicolumn{2}{c|}{0.86} & \multicolumn{2}{c|}{0.60} & \multicolumn{2}{c|}{0.18} & \multicolumn{2}{c|}{0.67} & \multicolumn{2}{c|}{\textbf{0.61}} & \multicolumn{2}{c|}{0.56} & \multicolumn{2}{c|}{0.44} & \multicolumn{2}{c|}{\textbf{0.81}} & \multicolumn{2}{c|}{0.70} & \multicolumn{2}{c}{0.70} \\ \hline
\multirow{2}{*}{\textbf{\begin{tabular}[c]{@{}c@{}}SURF\\ +OCR\end{tabular}}} & 0.54 & 1.00 & 0.39 & 0.88 & 0.75 & 0.60 & 0.83 & 0.50 & 0.50 & 0.73 & 0.60 & 0.64 & 0.20 & 0.89 & 0.47 & 1.00 & 0.53 & 0.82 & 0.65 & 0.83 \\ \cline{2-21} 
 & \multicolumn{2}{c|}{0.70} & \multicolumn{2}{c|}{0.54} & \multicolumn{2}{c|}{\textbf{0.67}} & \multicolumn{2}{c|}{0.63} & \multicolumn{2}{c|}{0.59} & \multicolumn{2}{c|}{0.62} & \multicolumn{2}{c|}{0.33} & \multicolumn{2}{c|}{0.64} & \multicolumn{2}{c|}{0.64} & \multicolumn{2}{c}{0.73} \\ \hline
\multirow{2}{*}{\textbf{\begin{tabular}[c]{@{}c@{}}SD+SURF\\ +OCR\end{tabular}}} & 0.78 & 1.00 & 0.47 & 1.00 & 0.58 & 0.70 & 0.83 & 0.50 & 0.50 & 0.73 & 0.65 & 0.79 & 0.40 & 0.67 & 0.65 & 0.94 & 0.75 & 0.82 & 0.73 & 0.92 \\ \cline{2-21} 
 & \multicolumn{2}{c|}{\textbf{0.88}} & \multicolumn{2}{c|}{\textbf{0.64}} & \multicolumn{2}{c|}{0.64} & \multicolumn{2}{c|}{0.63} & \multicolumn{2}{c|}{0.59} & \multicolumn{2}{c|}{\textbf{0.71}} & \multicolumn{2}{c|}{\textbf{0.50}} & \multicolumn{2}{c|}{0.77} & \multicolumn{2}{c|}{\textbf{0.78}} & \multicolumn{2}{c}{\textbf{0.81}}
 \end{tabular}
\caption{Aggregated precision, recall, and F1 scores of the state diagram generation process of StateLens using a combination of features -- Screen Detection (SD), SURF, and OCR -- and with stationary, hand-held, and web (with links) usage videos. Each cell shows the precision (top left), recall (top right), and F1 scores (bottom). Bold values identify the feature combination with the best performance.}
\label{tab:megatable}
\end{table*}

\section{Technical Evaluation}
We conducted a multi-part technical evaluation in order to understand how each key component of StateLens performs across a wide range of interfaces and usage scenarios.

\subsection{Dataset}
We collected a total of 28 videos from a diverse set of eight dynamic touchscreen interfaces, in different lighting conditions, and with both stationary and hand-held cameras, resulting in a total of 40,140 video frames. We also manually selected web videos of four touchscreen interfaces, resulting in a total of 32,610 video frames. All of these videos for our evaluation were collected by sighted people. The list of interfaces is shown in Figure~\ref{fig:interfaces}, and summarized in Table~\ref{tab:megatable}.

\subsection{Generating the State Diagram}
We first evaluated the effectiveness of StateLens in reconstructing interface structures from stationary, hand-held, and web usage videos. After StateLens generated the states, researchers manually coded them as correct, missing, or redundant in order to calculate precision and recall. A high precision indicates that most of the extracted states are unique screens of the actual interface (few duplicates). A high recall indicates that most of the screens of the interface are captured in the extracted states (good coverage).

For each interface and video source, we computed the precision, recall, and F1 scores for the extracted states using four configurations of features: {\em (i)} SURF features only, {\em (ii)} Screen Detection and SURF features, {\em (iii)} SURF and OCR features, and {\em (iv)} Screen Detection, SURF, and OCR features. The results are shown in Table~\ref{tab:megatable}. Overall, the combination of Screen Detection+SURF+OCR features achieved high performances across a wide range of interfaces, and were often the best in the four feature configurations.

Regarding the effect of our screen detection approach, a combination of Screen Detection+SURF+OCR features generally yielded higher performance compared to SURF+OCR features. The advantages were mostly observed in the precision differences and especially for web videos, as irrelevant frames and noisy background were filtered out. The screen detection technique did not work well for the Coca-Cola machine, as the object detection model would not classify it as electronics or machines. To address this problem, special-purpose models for detecting screens could be built.

Regarding OCR features, a combination of Screen Detection+SURF+OCR features generally had better performance compared to Screen Detection+SURF features. The advantages were mostly observed in the recall differences, and specifically for interfaces that had many similar screens in graphical layout with only text changes, {\em e.g.}, coffee machine (graphical), coffee machine (text-only), projector control, and room reservation interfaces. Regarding the different video sources, stationary videos generally performed better compared to hand-held ones for the same interface, because state matching is more robust with less camera blur, changing background noise and other uncertainty from camera motion.

Parameters can be chosen to further maximize recall (sacrificing some precision), as post-hoc crowd validation can be applied in the future to further filter out duplicates. Duplicate states require more manual effort to clean up, but have less impact on user experience compared to missing states.

\subsection{Accessing the State Diagram}
We next evaluated the effectiveness of using state diagrams to reduce latency and prevent errors in the state detection process.

\begin{figure}[b!]
 \centering
 \includegraphics[width=\linewidth]{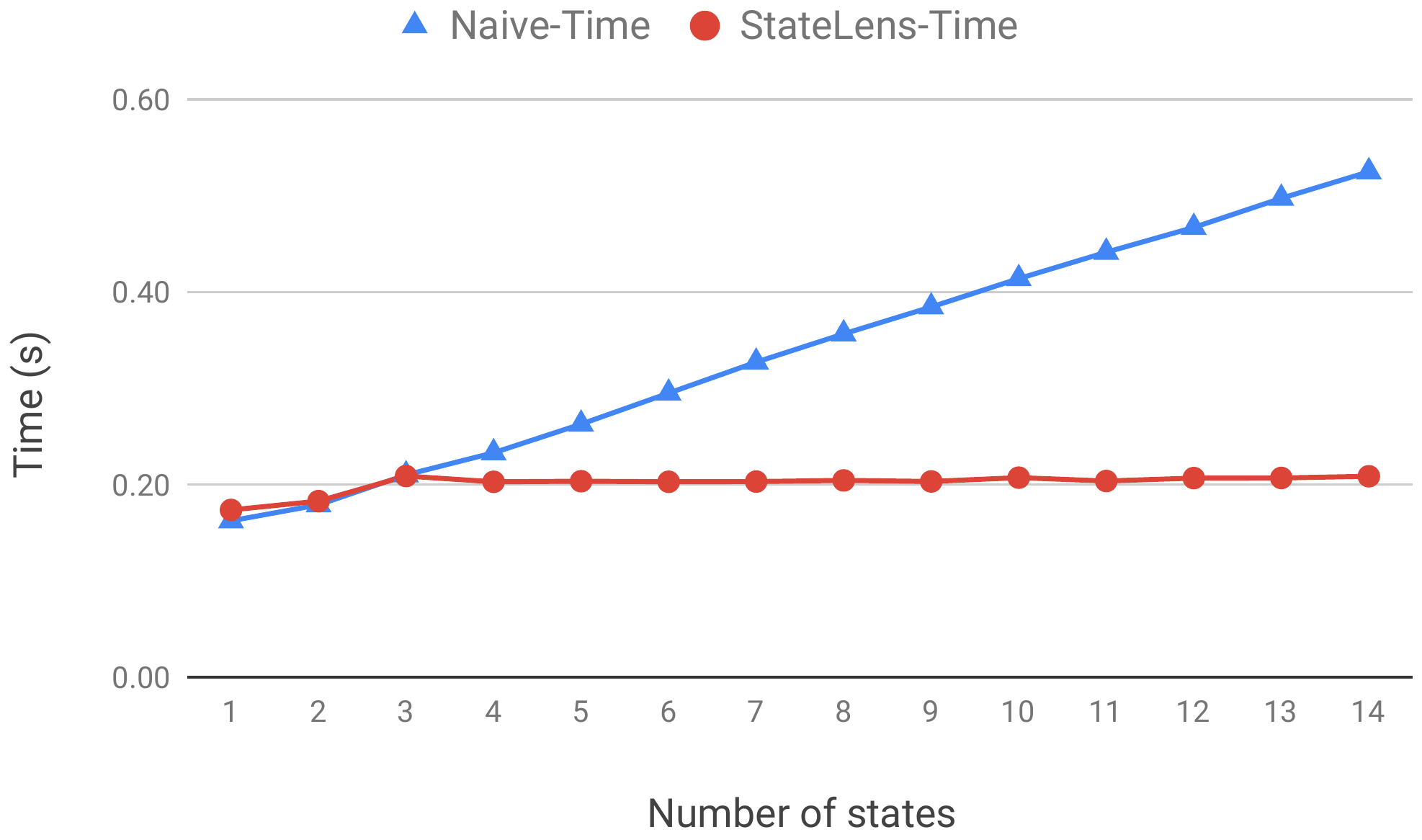}
 \vspace{-0.6pc}
 \caption{StateLens maintains a relatively stable processing time for state detection as the number of states increases, compared to the linear increase in the baseline approach.}
 \label{fig:scalability_time}
 \vspace{-0.1pc}
\end{figure}

\subsubsection{Using State Diagram to Reduce Search Time}
We evaluated the efficiency of our techniques in identifying states compared to the naive approach in VizLens::State Detection \cite{guo2016vizlens} which compares against every possible reference image. We varied the total number of states involved from one to all 14, and plotted the amount of processing time required for identifying the current state. The results show that as the number of states increases, StateLens achieved a relatively stable processing time compared to the linear increase in the baseline approach (Figure~\ref{fig:scalability_time}). Furthermore, using the coffee machine with all 14 states, StateLens can still maintain sufficient speed for audio-guided interaction (\textasciitilde5fps), while the baseline approach dropped to \textasciitilde2fps and became unusable.

\begin{figure}[t!]
 \centering
 \includegraphics[width=\linewidth]{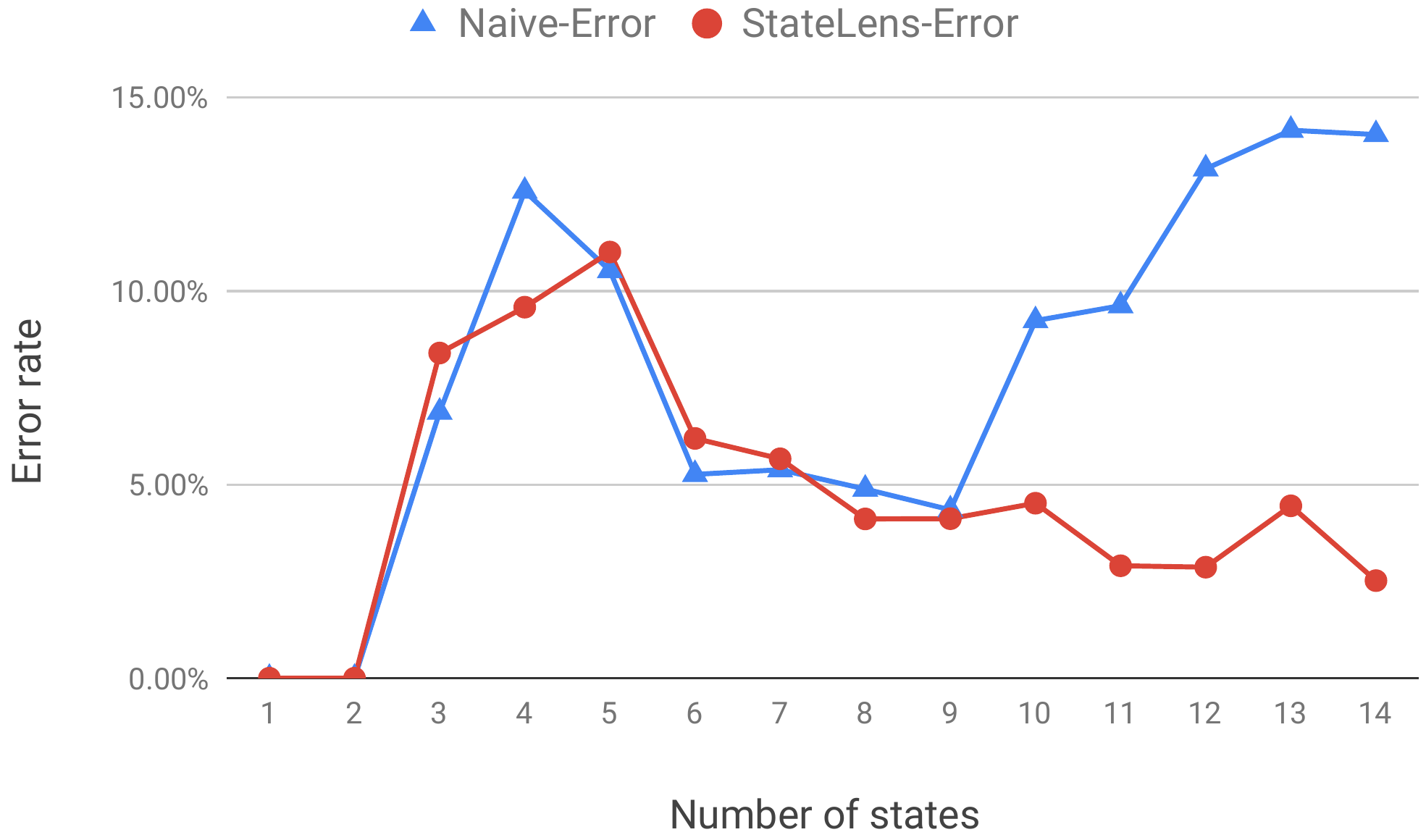}
 \vspace{-0.6pc}
 \caption{StateLens maintains a relatively stable error rate for state detection as the number of states increases, compared to the increasing trend in the baseline approach.}
 \label{fig:scalability_error}
 \vspace{-0.5pc}
\end{figure}

\subsubsection{Using State Diagram to Reduce Search Error}
We then evaluated the robustness of our techniques in identifying states compared to the baseline approach. We varied the total number of states involved from one to all 14, and plotted the percentage of errors in identifying the current state. The results show that as the number of states increases, StateLens achieved a relatively stable error rate of \textasciitilde5\% compared to the increasing trend in the baseline approach (Figure~\ref{fig:scalability_error}). Next in user evaluation, we further demonstrate how the generated state diagrams power interactive applications to assist blind users access existing dynamic touchscreen devices.

\begin{table*}
 \renewcommand{\arraystretch}{1.3} 
\centering
\footnotesize
\begin{tabular}{l|l|l|l|l|l|l}
\textbf{ID} & \textbf{Gender} & \textbf{Age} & \textbf{Occupation} & \textbf{Vision Level} & \textbf{Hearing} & \textbf{Smartphone Use} \\ \specialrule{.1em}{.05em}{.05em} 
P1 & Female & 64 & Retired & Light perception, since 10 years old & Normal & iPhone, 9 years\\ \hline
P2 & Female & 77 & Retired & Light perception & Normal & iPhone, 2 years\\ \hline
P3 & Female & 34 & Unemployed & Blind, since birth & Normal & iPhone, 6.5 years\\ \hline
P4 & Female & 46 & AT consultant & Blind, since birth & Normal & iPhone, 5 years \\ \hline
P5 & Male & 43 & IT consultant & Light/motion perception & Slight loss & iPhone, 3.5 years\\ \hline
P6 & Male & 67 & Business Rep. & Blind, since birth & Normal & iPhone, 5.5 years\\ \hline
P7 & Female & 64 & Retired & Blind, since birth & Mild loss & iPhone, 7.5 years\\ \hline
P8 & Male & 85 & Retired & Blind, since 8 years old & Normal & No \\ \hline
P9 & Female & 37 & AT Director & Light/shape perception & Normal & iPhone, 6 years\\ \hline
P10 & Female & 73 & Retired & Blind, since birth & Normal & iPhone, 2 years\\ \hline
P11 & Female & 71 & Retired & Blind, since childhood & Slight loss & iPhone, 7.5 years\\ \hline
P12 & Male & 71 & Retired & Low vision (20/200), color blind, since birth & Normal & iPhone, 9 years \\ \hline
P13 & Female & 51 & Unemployed & Blind, since birth & Moderate loss & iPhone, 8 years \\ \hline
P14 & Male & 71 & Retired & Light perception & Slight loss & iPhone, 4 years \\
\end{tabular}
\vspace{0.5em}
\caption{Participant demographics for our user evaluation with 14 visually impaired users. Thirteen were blind, and one (P12) had low vision.}
\vspace{-1pc}
\label{tab:maindemographics}
\end{table*}

\section{User Evaluation}
The goal of our user study was to evaluate how the components of StateLens (the 3D-printed accessories, the conversational agent, and the iOS application) perform in enabling blind people to accomplish realistic tasks that involve otherwise inaccessible dynamic touchscreen interfaces.

\subsection{Apparatus and Participants}
In order to enable repeated testing without wasting coffee, we built a simulated interactive prototype of the coffee machine in Figure~\ref{fig:CV} with InVision \cite{InVision}, which we displayed on an iPad tablet of similar size as the coffee machine's interface (iPad Pro 3rd generation, 11-inch, running iOS 12.2 without VoiceOver enabled). The conversational agent and the iOS application were installed on an iPhone 6, running iOS 12.2 with VoiceOver enabled. The finger cap and the conductive stylus in Figure~\ref{fig:accessories} were fabricated and used. We recruited 14 visually impaired users (9 female, 5 male, age 34-85). The demographics of our participants are shown in Table~\ref{tab:maindemographics}.

\subsection{Procedure}
Following a brief introduction of the study and demographic questions, participants first completed tasks using the 3D-printed accessories. For each of the three screen placements (in the order of \ang{90} vertical at chest-level, \ang{45} tilted at chest-level, and \ang{0} flat on the table), participants completed five trials using both the finger cap and the conductive stylus. The order of accessories was counterbalanced for all participants. For each trial, participants were first instructed to \textit{explore} by placing the accessory on the touchscreen and move according to the researcher's verbal instructions without activating touches. Participants were then asked to \textit{activate} a touch. The number of accidental triggers during exploration, and the number of attempts during activation were recorded. 

Next, participants were asked to talk to the conversational agent to prespecify drinks they want to order from the coffee machine for three times. Participants were instructed to order from a general category ({\em e.g.}, coffee drinks), but can freely choose the other properties ({\em e.g.}, coffee type, strength, size). Task completion rate and time were recorded.

Next, according to the three interaction traces prespecified through the conversational agent, participants were asked to use the 3D-printed accessories to perform the tasks following the guidance and feedback of the iOS application. These realistic tasks involved a series of button pushes across many states, {\em e.g.}, select gourmet drinks, cafe latte, strong strength, then confirm, auto-select default coffee bean, and end on the drink preparation screen. The iPad Pro simulating the inaccessible coffee machine was placed tilted at chest level, and the iPhone 6 running the iOS application was mounted on a head strap to simulate a head-mounted camera. Task completion rate and time were recorded.

After each step of the study, we collected Likert scale ratings and subjective feedback from the participants. Finally, we ended the study with a semi-structured interview asking for the participant's comments and suggestions on the StateLens system. The study took about two hours and participants were each compensated for \$50. The whole study was video and audio recorded for further analysis.

\subsection{Results}
We now detail our user study results and summarize user feedback and preferences. For all Likert scale questions, participants rated along a scale of 1 to 7, where 1 was extremely negative and 7 was extremely positive.

\begin{table}[b!]
\footnotesize
\centering
\renewcommand{\arraystretch}{1.5}
\begin{tabular}{l|l|l|l|l}
\hline
 \multicolumn{2}{l|}{\textbf{Screen Placement}} & \textbf{90 degrees} & \textbf{45 degrees} & \textbf{0 degree} \\ \hline
\multirow{6}{*}{\textbf{Stylus}} & Triggers & 0 (0) & 0.05 (0.21) & 0.03 (0.17) \\ \cline{2-5} 
 & Attempts & 2.63 (1.13) & 2.52 (1.08) & 2.29 (0.99) \\ \cline{2-5} 
 & Learnability & 6.0 (0.9) & 6.3 (0.9) & 6.3 (0.9) \\ \cline{2-5} 
 & Comfort & 5.8 (1.4) & 6.0 (1.1) & 6.1 (0.9) \\ \cline{2-5} 
 & Usefulness & 6.0 (1.2) & 6.3 (0.6) & 6.5 (0.7) \\ \cline{2-5} 
 & Satisfaction & 5.8 (1.3) & 6.5 (0.5) & 6.7 (0.5) \\ \hline
\multirow{6}{*}{\textbf{Cap}} & Triggers & 0.09 (0.34) & 0.06 (0.24) & 0.05 (0.21) \\ \cline{2-5} 
 & Attempts & 2.12 (1.10) & 1.75 (0.95) & 1.81 (0.96) \\ \cline{2-5} 
 & Learnability & 6.1 (0.6) & 6.2 (1.2) & 6.5 (0.7) \\ \cline{2-5} 
 & Comfort & 5.3 (1.4) & 6.5 (0.7) & 6.1 (0.9) \\ \cline{2-5} 
 & Usefulness & 6.3 (0.6) & 6.6 (0.7) & 6.5 (0.8) \\ \cline{2-5} 
 & Satisfaction & 6.2 (0.8) & 6.6 (0.7) & 6.5 (0.7) \\ \hline
\multicolumn{2}{l|}{\textbf{Preference (S/C)}} & 54\% / 46\% & 38\% / 62\% & 31\% / 69\% \\ \hline
\end{tabular}
\caption{Results from the 3D-printed accessory study, showing mean and standard deviation (in parentheses).}
\label{tab:touchstudy}
\end{table}

\subsubsection{Exploration and Activation with 3D-Printed Accessories}
All participants except P12 completed tasks using the 3D-printed accessories. P12 had low vision, and was able to hover his finger above the target and then activate by himself. The aggregated results are shown in Table~\ref{tab:touchstudy}. Using the conductive stylus to explore touchscreens generally resulted in fewer accidental triggers $(\textit{M} = 0.03, \textit{SD} = 0.16)$ compared to using the finger cap $(\textit{M} = 0.07, \textit{SD} = 0.27)$. On the other hand, the average attempts of using the stylus $(\textit{M} = 2.48, \textit{SD} = 1.07)$ was more than that from using the finger cap $(\textit{M} = 1.90, \textit{SD} = 1.01)$. This is likely because the conductive material is less sensitive compared to fingers.

In general, participants found both accessories to be comfortable to use $(\textit{M} = 5.9, \textit{SD} = 1.1)$ and highly useful $(\textit{M} = 6.4, \textit{SD} = 0.8)$. However, there were differences across the various screen placements. Participants slightly preferred using the stylus to explore and activate touchscreens in the \ang{90} screen placement (54\% vs. 46\%), since holding the hand in the upright position using the finger cap was not as comfortable $(\textit{M} = 5.3, \textit{SD} = 1.4)$, and the stylus felt more natural. Others preferred the finger cap since it provided better control over the stylus. On the other hand, participants preferred the finger cap much more than the stylus (65\% vs. 35\%) in the \ang{45} and \ang{0} screen placements, since the finger cap became more comfortable to use in these positions $(\textit{M} = 6.3, \textit{SD} = 0.8)$. 

We observed that participants sometimes held the accessories in awkward postures, likely due to unfamiliarity. This can be improved with practice, as participants generally found the accessories to be very easy to learn $(\textit{M} = 6.2, \textit{SD} = 0.9)$. Better affordances could further improve learnability as one participant (P14) noted that a conductive stylus design which incorporates a physical button to trigger, instead of a conductive region, would be beneficial. 

Another interesting observation was that 8 of 13 participants who completed the tasks for the printed accessories would occasionally perform a ``double-click'', or two taps in quick succession to activate the screen. Almost all of this subset of participants (7) had a strong familiarity with using VoiceOver on an iPhone or iPad, suggesting their habitual use of this technology may influence their interactions using the accessories.

\subsubsection{Prespecifying Tasks with the Conversational Agent}
Participants spent an average of 53.7 seconds $(\textit{SD} = 11.6)$ to prespecify tasks with the conversational agent, with an overall task completion rate of 100\%, and found it to be extremely easy to learn $(\textit{M} = 6.6, \textit{SD} = 0.6)$, comfortable to use $(\textit{M} = 6.8, \textit{SD} = 0.4)$, and useful $(\textit{M} = 6.7, \textit{SD} = 0.6)$. Several participants tried specifying multiple parameters in one sentence ({\em e.g.}, I want a large coffee 50-50, shown in Figure~\ref{fig:agentinteraction}). Note that the task completion time is likely to reduce in practice, since the agent's speaking rate is dependent on the users' screen reader setting, and after repeated usage, the users will get familiar with the functions.

\subsubsection{Completing Realistic Tasks}
Participants spent an average of 122.3 seconds $(\textit{SD} = 41.9)$ completing the first task, 110.4 seconds $(\textit{SD} = 36.9)$ for the second, the 97.6 seconds $(\textit{SD} = 30.7)$ for the third, as they got familiar with the audio feedback and guidance. The overall task completion rate was 94.7\%. For five of the tasks, participants accidentally selected the wrong option and had to go back or start over. Because our smoothing approach requires a new state to be seen continuously across multiple frames in order to determine a state transition, there may be a delay in determining if a button press was successful. In this case, some users may accidentally press again at the same location triggering an incorrect selection on the next screen state. This issue may be alleviated by providing more immediate feedback such as a tentative audio confirmation that a button press has been successful.

In subjective ratings, participants found the StateLens iOS application to be easy to learn $(\textit{M} = 5.5, \textit{SD} = 0.9)$, comfortable to use $(\textit{M} = 5.6, \textit{SD} = 1.2)$, and very useful $(\textit{M} = 6.1, \textit{SD} = 1.1)$. They felt the audio feedback provided by the app was in real-time and accurate $(\textit{M} = 6.1, \textit{SD} = 0.9)$. Participants mentioned that the head mount can be made more comfortable using a lighter setup, {\em e.g.}, glasses.

Overall, participants were very excited about the potential of StateLens, and felt that it could help them access other inaccessible interface in the future $(\textit{M} = 6.6, \textit{SD} = 0.9)$:

\begin{quote}
 {\em ``It'll be a thing, I will actually use it.''} -- P1
\end{quote} 

\begin{quote} 
 {\em ``[StateLens] gives much more flexibility, so that if the machine itself doesn't have speech, this can cover the instances where you have to interact with a touchscreen. There are more tools to access them. This combination opens up more accessibility. ... I can't wait to see this in action!''} -- P6
\end{quote} 

\begin{quote}
 {\em ``I really like the idea of using the phone to make screens accessible and give feedback in real time. That's really impressive. I would use it. It would be helpful and useful.''} -- P9
\end{quote} 

\begin{quote}
 {\em ``I would welcome more opportunities to use interfaces with [StateLens], like operating the cable company box. It would be great if interfaces could also show up on my phone screen and read it to me or let me explore it there.''} -- P12
\end{quote}

A low vision user (P12) mentioned that even though he might not always need assistance, if the interface's contrast or brightness is poor, a system like StateLens would be greatly helpful as a confirmation. Furthermore, he would like to get more information beyond the text labels on the buttons by using StateLens as a cognitive assistant. He would find it useful if, for example, a button for a coffee selection labeled ``Rainbow's End'' could further be described as ``a coffee blend containing tasting notes of nuts and citrus'' even though the display does not provide that information.

\section{Discussion and Future Work}
In this section, we discuss how the approaches used in StateLens might generalize to extract information from existing online videos to, for instance, assist sighted users and construct a queryable map of devices. We also discuss limitations of our work, which represent opportunities for future research.

\subsection{Technical Approach to Accessibility}
StateLens is not the ideal solution. In a perfect world, post-hoc fixes like StateLens would not be needed (because all technologies would be inherently accessible), but in practice access technology like StateLens plays a vital role. Even with the existing laws, there are still many cases where ``reasonable accommodation'' is not enough. For example, a vending machine could be labeled with Braille, but the checkout credit card machine is not accessible. StateLens is a stopgap measure to make access possible (as are many access technologies), and introduces ideas that might find purchase in other access and accessible technologies.

People who are blind were involved throughout the research, including several people with visual impairments on our extended research team, and multiple sessions of design and study with a total of 30 outside participants. While we strove to make this paper self-contained, it builds on our long history of work involving thousands of blind people as students, researchers, participants, and users.

\subsection{Generalizability}
In this paper, we developed a hybrid crowd-computer vision system to enable access to dynamic touchscreens in-the-wild. One unique contribution of this work is that we demonstrated the possibility of extracting state diagrams from existing point-of-view videos instead of screenshots or screencast videos \cite{banovic2012waken, lafreniere2013community, wang2018workflow}. For existing physical devices whose underlying hardware or software cannot be modified, point-of-view videos are more prevalent and easier to acquire compared to screencast videos, which makes our approach generalizable to a large variety of devices and scenarios.

We motivated our approach as a benefit to improve accessibility for blind users. However, this approach could be beneficial to sighted people and people with cognitive disabilities in many ways as well. For example, medical devices can be hard to configure, and devices that are in foreign languages are hard to operate. Through understanding of the state diagrams of devices with readily available or user-taken point-of-view videos, our approach can provide additional information to the user as they interact with the devices ({\em e.g.}, augmented reality applications for translation services, interactive tutorials).

Using StateLens, we envision building a queryable map of state diagrams for many of the devices in the world using existing point-of-view videos that have been shared online. As users start to use a device, it can be geo-located, automatically recognized, or added into the system. Additional states can be added to the existing diagram as users interact with the device. Changes to the devices can be automatically detected over time to update the interface state diagram. Furthermore, similar but slightly different models of a device may reuse another state diagram and enable transfer learning.

\subsection{Assistive Hardware for Automatic Screen Actuation}
Our 3D-printed accessories elegantly add ``risk-free exploration'' to existing capacitive touchscreen devices without modifying the underlying hardware or software, which has been a major hurdle for past efforts. In our user study, we discovered issues around holding the accessories in certain angles, and ``the last meter'' problem to accurately activate the exact button once. If the screen is cluttered, it could still be quite difficult to operate. As future work, we have started to design hardware proxies that can locate and actuate external touchscreens automatically. Blind users could brush a ``phone case'' on the external touchscreen, then the built-in camera would capture, recognize, and instruct actuators contacting the external screen to trigger functions at the right place and time.

\subsection{Limitations}
As with most systems, StateLens currently has some limitations, which we believe could be explored in future work. For instance, StateLens has limited capability in noticing and differentiating minor interface changes such as toggle buttons or color indicators. One solution may be to detect and factor in UI widgets that are expected to change using approaches like those in PreFab \cite{dixon2010prefab} and TapShoe \cite{swearngin2019modeling}. Furthermore, StateLens cannot currently handle major updates and layout changes of the interface, as well as list menus, slide bars or other gestures ({\em e.g.}, scroll, swipe, pinch).

The completeness of the state diagram is limited by the coverage of the videos collected for the device. Even if videos only capture a subset of possible tasks, these would likely be frequently used paths of action, thus still providing reasonable functionality in many cases. If a blind user needs to access an unseen state, StateLens could add it to the state diagram on-the-fly, asking the user to wait for that screen to be labeled and then added to the full state diagram. Other approaches include generalizing based on the existing states or other machines, and relying more on OCR.

We evaluated StateLens across a number of touchscreen interfaces and with blind users in the lab, but we did not deeply study how StateLens works in the real world, which is often much more complicated and messier than in-lab studies. Our next step is to harden our implementation to scale to many users, and deploy it to understand how it performs in the everyday lives of blind people.

\section{Conclusion}
We have presented {\em StateLens}, a reverse engineering solution that makes existing dynamic touchscreens accessible. Using a hybrid crowd-computer vision pipeline, StateLens generates state diagrams about interface structures from point-of-view usage videos. Through these state diagrams, StateLens provides interactive feedback and guidance to help blind users prespecify task and access the touchscreen interfaces. A set of 3D-printed accessories enable capacitive touchscreens to be used non-visually by preventing accidental touches on the interface. Our formative study identified challenges and requirements, which informed the design and architecture of StateLens. Our evaluations demonstrated the feasibility of StateLens in accurately reconstructing the state diagram, identifying interface states, and giving effective feedback and guidance. More generally, StateLens demonstrates the value in a hybrid, reciprocal relationship between humans and AI to collaboratively solve real-world, real-time accessibility problems.

\section{Acknowledgments}
This work has been supported by the National Science Foundation (\#IIS-1816012), Google, and the National Institute on Disability, Independent Living, and Rehabilitation Research (NIDILRR). We thank the participants who contributed to our studies for their time, and the reviewers for their valuable feedback and suggestions. Special thanks to Patrick Carrington, Meredith Ringel Morris, Zheng Yao, and Xu Wang for their help and support.

\balance{}

\balance{}

\bibliographystyle{SIGCHI-Reference-Format}
\bibliography{main}

\end{document}